\documentstyle[prl,aps,preprint]{revtex}

\begin{document}
\draft

\title{
Resonant Tunneling Between Quantum Hall Edge States
}

\author{K. Moon[1], H. Yi[2], C.L. Kane[2], S.M. Girvin[1],
Matthew P.A. Fisher[3]}

\address{ {\rm[1]}
Department of Physics, Indiana University,
Bloomington, Indiana 47405
}

\address{ {\rm [2]}
Department of Physics, University of Pennsylvania,
Philadelphia, PA 19104
}

\address{ {\rm [3]}
IBM Research, T.J. Watson Research Cent., PO Box 218,
Yorktown Heights, NY 10598
}

\date{\today}
\maketitle

{\tightenlines
\begin{abstract}
Resonant tunneling between fractional quantum Hall edge states is
studied in the Luttinger liquid picture. For the Laughlin parent
states, the resonance line
shape is a universal function whose width scales to zero at zero
temperature.  Extensive quantum
Monte Carlo simulations are presented for $\nu = 1/3$ which
confirm this picture and provide a parameter-free prediction for the line
shape.
\end{abstract}
}


\narrowtext
Many body correlations often play an important role in tunneling
and resonant tunneling in mesoscopic structures, such as quantum
dots\cite{averin,meirwingreenlee}.
In addition to the Coulomb correlations in the vicinity
of a tunneling structure,
it has recently been emphasized that the electron
interactions in the ``leads'' feeding the structure can be crucial.
Specifically, when the leads are one-dimensional,
interactions are believed to de-stablize the Fermi liquid,
forming a Luttinger liquid.  In this case, the tunneling
conductance
through a point contact is predicted to vanish
as a power law of temperature, due to
the absence of long-lived single-particle excitations
in the incident electron gas\cite{kanefisherprl}.
Unfortunately, in real one-dimensional wires
extraneous impurities away from the point contact will
complicate matters,
tending to backscatter and localize the electrons.

In this article, we consider the situation where the ``leads''
feeding a point contact are 2d fractional quantum Hall fluids.
We argue that the correlations built into Laughlin states
play a crucial role in determining the nature of resonant tunneling
at low temperatures.  In particular, we show that as the
temperature
is lowered, resonances become sharper and have a universal shape.
We present the results of extensive quantum
Monte Carlo simulations which give the universal line shape expected for
the $\nu =1/3$ state.

Provided the temperature is below the gap in the fractional
quantum Hall fluid, the incoming current will be carried
by edge states.
In a beautiful series of articles, Wen\cite{WEN}
 has demonstrated that the gapless
edge excitations of a quantum Hall system are ``chiral'' Luttinger
liquids.  Since the electrons in a chiral Luttinger liquid
move only in one direction, backscattering
is only possible when opposite edges of the sample are close
together, namely at the point contact.  Thus localization
in the edge state ``leads'' is completely unimportant.

Recently, two of us have studied the problem of resonant tunneling in
a Luttinger liquid\cite{kaneandfisher}.
A renormalization group analysis reveals that
the resonance lineshape at low temperatures is universal,
depending only on the two terminal conductance,
$G = g e^2/h$ of the Luttinger liquid.  The
parameter $g$ depends in a complicated way
on the strength of the interactions between left- and right-moving
electrons.  In a quantum Hall edge state
the electrons move only in one direction.  Most strikingly,
in this case $g$ is a topological invariant controlled by the quantum
Hall state in the bulk\cite{WEN}.
To see this, consider raising the chemical
potential of the right-movers relative to that of the left by an amount
$\delta \mu$.  This
corresponds to applying a Hall voltage $V_{\rm H} = \delta\mu/e$
and the resulting current is
given by the quantized Hall coefficient as $I = \nu (e^2/h) V_{\rm H}$.
This immediately establishes the universal result
within a given Hall plateau:
$g = \nu$.
{\it This remarkable fact makes the resonance line shape completely
universal, model-independent
and fully determined (up to an overall temperature scale)\/}.  The
fractional quantum  Hall regime ($g<1$) is thus a far more promising
place to observe pure Luttinger liquid behavior than in a single-channel
quantum wire, where it is difficult to eliminate disorder and
where the value of $g$ is unknown.  Chamon and Wen\cite{chamonandwen}
have recently considered a theory of resonant tunneling between quantum
Hall edge states which is valid in the limit in which the peak conductance of
the resonances are much less than $e^2/h$, or in the tails of a
stronger resonance.  A similar theory for resonant tunneling in
a Luttinger liquid has been developed by Furusaki and
Nagaosa\cite{furusakinagaosa}.  In contrast, the scaling
theory presented below is valid over the entire width of the resonance
for resonances whose peak conductance
approaches the `perfect' value $\nu e^2/h$.

The analog of a weak impurity that causes back-scattering is a narrow
constriction which brings the left- and right-movers
close enough
together to communicate via tunneling of Laughlin quasiparticles
through a `weak link' as illustrated in Fig.(\ref{fig1}).
This can be achieved by means of a gate which electrostatically defines
a narrow region in the Hall bar.  The analog of the two-impurity
resonance geometry considered by Kane and Fisher\cite{kaneandfisher}
would be two nearby, symmetric tunneling paths
for quasiparticles\cite{geometry}.  For some value of
magnetic field or gate voltage one will (randomly or intentionally)
achieve the condition of destructive interference\cite{interference}
which shuts off the inter-edge
quasiparticle tunneling.  This is the resonance
(no backscattering) condition which will be manifested experimentally
by the appearance of a
two-terminal source-drain conductance\cite{fourterm}
which peaks at a value which at low temperatures approaches
the quantized value,
$G = \nu e^2/h$.  Away from
resonance the quasi-particle tunneling
causes current to leak
from one edge to the other, thereby reducing the
source to drain conductance (see
Fig.(\ref{fig1}).  In fact, in the fractional quantum Hall effect
the quasiparticle tunneling
is expected to increase upon cooling,
driving the conductance all the way to zero in the zero temperature
limit.

Following the analysis of ref. \onlinecite{kaneandfisher}, the weak
quasiparticle tunneling regime can be analyzed perturbatively.  As the
temperature is lowered and the effective tunneling amplitude grows,
perturbative analysis fails.  However at low enough temperatures
the conductance from
source to drain becomes tiny
and a perturbative analysis in terms of
electron transmission becomes possible.  The peak and tail regimes of
the resonance are tied together by a universal scaling function.  Below
we establish a model and calculate the scaling function
for $g=1/3$ using quantum Monte Carlo.

We begin our analysis
by briefly reviewing the logic behind Wen's edge state theory.
For simplicity we focus here on the primary Hall states
with inverse filling factor $\nu^{-1}$ equal to an odd integer.
In this case the edge state has only one branch.

Conservation of electron three-current $j_\mu$
permits us to introduce a fictitious gauge field $a_\mu$ via
\begin{equation}
j_\mu = {1 \over {2 \pi}} \epsilon_{\mu \nu \lambda}
\partial_\nu a_\lambda   .
\label{2}
\end{equation}
The bulk 2D electron gas is in an incompressible quantum Hall state
with an excitation gap, which means that the low-energy,
long-length-scale physics must be described by a massive theory.  In 2+1-D
the only massive gauge theory is the Chern-Simons theory which has
(Euclidean) action\cite{zee} (ignoring irrelevant terms):
\begin{equation}
S_{bulk} = {i \over {4 \pi \nu }} \int a_\mu \partial_\nu a_\lambda
\epsilon_{\mu \nu \lambda} d^2 x d \tau  ,
\label{3}
\end{equation}
The coefficient $\nu^{-1}$ is uniquely fixed by the quantized
Hall conductivity and
specifies the number of zeros bound to the electrons in the Laughlin wave
function\cite{smgandmacd,kivelsonzhang,leeandfisher}

Wen has shown that in the presence of a boundary, say at y=0,
an effective action for the edge state can be obtained
as follows\cite{WEN}:  First
integrate out $a_\tau$ in the bulk,
which gives an incompressibility constraint on the electron density,
$\epsilon_{ij} \partial_i a_j =0$.   Then solve the constraint
in terms of a scalar field,
$a_j = \partial_j \phi$.  After an integration by parts
the final Euclidean action for the
edge state takes the form,
\begin{equation}
S_{\rm edge} = {1 \over {4 \pi g}} \int dx d \tau (\partial_x \phi)
( i \partial_\tau \phi + v \partial_x \phi  )   .
\label{4}
\end{equation}
Here $v$ is the velocity of the edge excitation, which is non-universal,
and
will depend on the details of the edge confining potential and the
Coulomb interaction at the edge.
The dimensionless parameter $g$ on the other
hand is {\it universal} and depends only on the quantum Hall state,
$|g| = \nu$.
As emphasized by Wen the requirement that the Hamiltonian
associated with eq.(\ref{3})
be bounded below requires that $v/g$ be positive.
This analysis neglects the effects of a long range Coulomb
interaction which are expected to become relevant at very low temperatures
(roughly $T<10mK$ for a Hall bar of length $100 \mu{\rm m}$ and width
$1\mu{\rm m}$)  \cite{WEN,Glazman}.

It follows from eq.(\ref{2}) that the one-dimensional
electron density along the edge, $\rho (x)$,
is given by $\rho = \partial_x \phi /2 \pi$.  An expression for
the electron creation operator at the edge can be obtained by
combining this with the fact that the momentum operator
conjugate to $\phi$ is  $\Pi_\phi = \rho /g$.
Since adding an extra electron to the edge is equivalent
to creating an `instanton' in
$\phi$, in which $\phi$ changes by $2 \pi$ in the region
near $x$, the electron creation operator at the edge is simply
\begin{equation}
\psi (x) \sim \exp\left[2\pi i \int^x \Pi_\phi (x') dx'\right]
 = e^{i \phi(x)/g}  .
\label{5}
\end{equation}
A `vortex' or Laughlin quasiparticle at the edge is created simply by
$e^{i \phi(x)}$, which carries fractional charge $ge$.

We now suppose that the right-moving and left-moving edges are
coupled via a tunneling term, say at $x=0$.  The total action will
then have the form,
$S_L[\phi_L] + S_R[\phi_R] + V(\phi_L,\phi_R)$.
Lacking a specific model
we can not say whether quasiparticles or
electrons will tunnel more easily.  Presumably
the fractionally-charged
quasiparticles see a lower barrier, but the matrix elements
may compensate.  Instead, we
write down the most general form allowed by symmetry.  Taking the
weak link region to be at $x=0$, we have
\begin{equation}
V = \sum_{m=1}^\infty v_m \exp\left[im(\phi_L(x=0)-\phi_R(x=0))\right]
+ {\rm c.c.},
\label{7}
\end{equation}
where the $v_m$ are (complex) tunneling amplitudes.
The term $m=1$ represents the combined amplitudes for a quasi-electron
to tunnel from one edge to the other {\it or\/} a quasi-hole to tunnel
in the opposite direction.  These physically distinct processes lead to
the same final state and hence add coherently to produce $v_1$.  The
term $m=1/g$ corresponds to electron tunneling.  We have no {\it a
priori\/} knowledge of the $v_m$.  Fortunately, for $g=1/3$, all terms
except $v_1$ are irrelevant, having a negative renormalization group
eigenvalue, $1-gm^2$,
\cite{kaneandfisher}.
Thus at low enough
temperatures ($T\ll\Delta$, where $\Delta \sim 1 {\rm K}$ is the bulk
excitation gap) and small enough $v_1$, the irrelevant variables $v_m,
m> 1$, will flow to zero before $v_{\rm eff} \sim v_1/T^{1-g}$  has
grown large.  Thus the RG flow will follow a {\it universal\/}
trajectory away from the resonance fixed point ($v_1=0$) into the
insulating fixed point.

At finite temperature the renormalization group flows will be cut
off, and the system will end up somewhere along that universal
trajectory.  From this it follows that in the limit of
low temperature, the conductance as a function of the
resonance tuning parameter $\delta\sim |v_1|$ and the temperature
will obey the scaling form,
\begin{equation}
G(T,\delta) = \tilde G_g(c\delta/T^{1-g}),
\label{8}
\end{equation}
The scaling function $\tilde G_g(X)$ is {\it universal} in the
sense that it does not depend on microscopic details, but
is a property of the universal
trajectory connecting the two fixed points.  Since $g=\nu$ in
the quantum Hall effect, $\tilde G_g$ is completely determined by
the theory.  The parameter $c$ is a non-universal dimensionful
factor which sets the temperature scale.
By demanding that the scaling form (6) matches onto
the off-resonance conductance, which vanishes
as $G(T) \sim T^{2/g -2}$, implies
that the tails
of the scaling function should decay like $X^{-2/g}$, or
$X^{-6}$ for $g=1/3$.

Though the general properties of the scaling function, such as
the temperature dependence of the width and the exponent in the
tails are known, the detailed shape of the scaling function has
been calculated analytically\cite{kaneandfisher} only for $g=1/2$.
This problem
is idealy suited to Monte Carlo simulation, and we have
explicitly computed $\tilde G(X)$, verified the predicted
scaling behavior, and determined the entire scaling function.
Following ref. \onlinecite{kaneandfisher} we note that the
action is gaussian in $\phi(x)$ for $x\ne 0$ and so we integrate
out all degrees of freedom except
$\phi(\tau)\equiv \phi_L(x=0,\tau)-\phi_R(x=0,\tau)$.
This gives the action,
\begin{equation}
S =  {1\over {4\pi |g|}} \sum_{i\omega_n} |\omega_n| |\phi(\omega_n)|^2
   + v_1 \int_0^\beta  d\tau \cos \phi(\tau),
\label{9}
\end{equation}
where we have retained only the single relevant operator.
We have computed the finite frequency `two-terminal' conductance
from the Kubo formula,
\begin{equation}
G(\omega_n) = {e^2\over 2\pi h} |\omega_n| < |\phi(\omega_n)|^2 >.
\label{10}
\end{equation}
A hard cutoff $\Lambda$ is introduced by keeping only a finite number of
Matsubara frequencies L (typically $L <100$).  We also simulated a
dual version of the model in which the tunneling events are represented
by a plasma of logarithmically interacting `charges'\cite{kaneandfisher}.
Essentially identically results were obtained in the two approaches.

In order to extract information about the
temperature dependence of the D.C. conductance, analytic continuation
to zero frequency is necessary.  Though difficult to do
exactly, this may be done with sufficient accuracy by
fitting the finite frequency data to a rational function
[2/3] Pad\'e form in order to extrapolate to $\omega=0$.

In order to check the
code and to test our analytic continuation procedure, we
compare our results to an exact solution which is available
for $g=1/2$ in Fig.(\ref{fig2})  The solid line is the exact solution
derived in ref. \onlinecite{kaneandfisher}, and the data points correspond to
Monte Carlo simulations of $G$ as a function of
$cv_1/T^{1-g}$
for different system sizes corresponding
to $T=\Lambda/(31\pi)$ and $T=\Lambda/(41\pi)$.
$c$ is the
single non-universal parameter which is adjusted to obtain the fit.

In Fig.(\ref{fig3})
we display the results of our Monte Carlo simulation for
$g=1/3$.  In this case the tails of the resonance are predicted to
decay much faster, like $X^{-6}$.  The data
clearly scales.  If the coefficient $c$ in eq.(\ref{9}) is chosen
such that the scaling function varies as $\tilde
G(X) = g(1-X^2)$ for small
$X$, this simulation allows us to determine that for large $X$
$\tilde G(X) = K X^{-6}$ with $K = 2.6\pm 0.2$.

It should be emphasized that this scaling behavior is to be expected
for the fractional quantum Hall effect $\nu = 1/3$, but not
for the integer effect $\nu=1$ or higher-order fractions.
In the integer case the edge state is equivalent
to a non-interacting Fermi-liquid, and at low temperatures
the resonances should be temperature-independent and Lorentzian.
For all
higher-order fractions in Laughlin's sequence, $1/\nu$ an
odd integer, multi-quasiparticle backscattering processes,
$v_m$ in (5) for $m>1$, are also relevant and grow
at low temperatures.  Thus in these cases
the conductance at the peak of the ``resonance'' ($v_1=0$)
will decrease upon cooling, eventually
killing completely the resonance in the zero temperature
limit.  The higher order hierarchical
quantum Hall fluids,
such as $\nu=2/3,2/5,2/7,...$, have more than one branch of edge
states\cite{WEN,macdonald}, which complicates the analysis.
Resonant tunneling between hierarchical edges will
be considered in a subsequent paper.

\acknowledgements{
It is a pleasure to thank V. Chandrasaekar,
 A.~H. MacDonald, F. Milliken, H. Mooij, Mats Wallin and R. Webb
for thoughtful discussions.   
M.P.A.F. is grateful to the Institute for Theoretical Physics
in Santa Barbara where part of this work was carried out.
The work at Indiana University was supported by NSF DMR-9113911.
The work at Pennsylvania was supported by the NSF MRL program,
under grant no. DMR91-20668.

}

\begin{figure}
\caption{Four terminal Hall bar geometry with a narrow constriction between
the source (S) and drain (D) formed by a lithographically patterned gate
(G).  The dotted lines represent two parallel
tunneling paths for quasiparticles.  Precisely on resonance, there is
perfect edge transmission between the source (S) and the drain (D) as
indicated by the arrows.  Away from resonance the edge channels are
reflected at low temperatures.}
\label{fig1}
\end{figure}

\begin{figure}
\caption{Log-log plot comparing resonance scaling function for $g=1/2$
obtained by
Monte Carlo simulation for two different temperatures (squares and hexagons)
and the exact solution (solid line).
The dashed line is the asymptotic behavior predicted to decay as $X^{-4}$.}
\label{fig2}
\end{figure}

\begin{figure}
\caption{Log-log plot of resonance scaling function for $\nu=1/3$ fractional
quantum Hall effect obtained by Monte Carlo simulation for two different
temperatures.  The dashed line is the asymptotic behavior predicted to decay
as $X^{-6}$.}
\label{fig3}
\end{figure}


\begin{references}
\bibitem{geometry}  These  tunneling paths (weak links)
may be due to random impurity potentials or an intentionally created
quantum dot.  If the tunneling occurs through a dot with significant
charging energy U, or the tunneling matrix element between the
dot from the leads is very small, a more careful analysis
is required in order to see how the resonance fixed point
is reached.  This has been studied in detail in
ref. \onlinecite{kaneandfisher}.
\bibitem{interference}  Since the Laughlin quasiparticles are vortices,
the relative phase for the tunneling on the two weak links is given by
the number of enclosed electrons in the area between the two links.
This is adjustable by means of the B field and possibly
electrostatically.
\end{references}
\end{document}